\documentclass[11pt,a4paper]{article}
\usepackage[utf8]{inputenc}
\usepackage{amsmath,amsthm,amsfonts,amssymb}
\usepackage{graphicx}
\usepackage{hyperref}
\usepackage{cite,float}
\usepackage{multicol}
\usepackage{authblk}
\hypersetup{colorlinks=true, linkcolor=blue, citecolor=blue, filecolor=blue, urlcolor=blue}

\usepackage{caption}
\usepackage[margin=1.1in]{geometry}
\usepackage{color}
\usepackage{verbatim}

\usepackage{booktabs}
\usepackage[svgnames,table]{xcolor}
\usepackage[tableposition=above]{caption}
\usepackage{pifont}

\usepackage[labelfont=bf]{caption}

\title{Quantum Cosmology of Fab Four John Theory with Conformable Fractional Derivative}
\author[1]{Isaac Torres\footnote{itsufpa@gmail.com}}
\author[2,3]{J\'ulio C. Fabris}
\author[2,4]{Oliver F. Piattella}
\author[2]{Ant\^onio B. Batista$^{\dagger}$}
\affil[1]{\small PPGFis, CCE - Universidade Federal do Esp\'{i}rito Santo, 29075-910, Vit\'{o}ria, ES, Brazil}
\affil[2]{N\'{u}cleo Cosmo-UFES \& Departamento de F\'{i}sica - Universidade Federal do Esp\'{i}rito Santo 29075-910, Vit\'{o}ria, ES, Brazil}
\affil[3]{National Research Nuclear University MEPhI, Kashirskoe sh. 31, Moscow 115409, Russia}
\affil[4]{Institut f\"ur Theoretische Physik, Ruprecht-Karls-Universit\"at Heidelberg, Philosophenweg 16, 69120 Heidelberg, Germany}
\date{}

\begin{document}
\maketitle
\vspace{-1.5cm}

\begin{center}
	\begin{minipage}{13.5cm}
		\small $^{\dagger}$ Prof. Ant\^onio Brasil Batista, one of the co-authors of this article, died the 01/03/2020, at the age of 81 years old. Very active, giving regularly lectures at the PhD course of Physics at UFES (Brazil), participating in the orientation of Master and PhD students, he was interested in the recent times, among other topics, in the applications and development of fractional derivatives, in particular in quantum cosmological systems, which is the subject of the present article, his last scientific contribution.
	\end{minipage}
\end{center}

\begin{abstract}
We study a quantization via fractional derivative of a nonminimal derivative coupling cosmological theory, namely, the Fab Four John theory. Its Hamiltonian version presents the issue of fractional powers in the momenta. That problem is solved here by the application of the so-called conformable fractional derivative. This leads to a Wheeler-DeWitt equation of second order, showing that a Bohm-de Broglie interpretation can be constructed. That combination of fractional quantization and Bohmian interpretation provides us a new quantization method, in which the quantum potential is the criterion to say if a quantum solution is acceptable or not to be further studied. We show that a wide range of solutions for the scale factor is possible. Among all of those, a bouncing solution analogous to the perfect fluid cosmology seems to deserve special attention.
\end{abstract}



\section{Introduction}
The expansion of the universe is one of the greatest discoveries of the last century. One of the more intriguing questions raised up by that discovery is how to describe the primordial universe at Planck's scales. It is a common idea that such a scenario may be better understood in terms of a quantum theory of cosmology \cite{bojowald2011quantum}. In this sense, the scalar fields have shown to be an insightful way to modify gravity \cite{Papantonopoulos:2015cva,Capozziello:2010zz}, by the introduction of an extra degree of freedom, that seems to be substantial for the early universe, since the canonical scalar field is the basis for inflationary models \cite{liddle2000cosmological,Chervon2019}. Some of those models deal with a fine-tuned potential, which is considered by some authors as a problem in inflationary models (see \cite{PhysRevD.80.103505}, for instance). Thus, it is important to look for generalizations and alternatives to the canonical scalar field coupling to gravity, known as minimal coupling. Among all possibilities, the Horndeski theory \cite{Horndeski:1974wa} is the most  general scalar-tensor theory in four spacetime dimensions with second-order equations of motion. It is represented by the action
\begin{equation}\label{horndact}
S=\int d^4x\sqrt{-g}(L_2+L_3+L_4+L_5)\; ,
\end{equation}
where
\begin{align}\label{hrndskip}
	L_2 & =G_2(\phi,X)\; ,  \\
	L_3 & =-G_3(\phi,X)\Box\phi\; ,  \\
	L_4 & = G_4(\phi,X)R+G_{4,X}(\phi,X)[(\Box\phi)^2  -\nabla^{\mu}\nabla^{\nu}\phi\nabla_{\mu}\nabla_{\nu}\phi]\; ,\\
	L_5 & = G_5(\phi,X)G^{\mu\nu}\nabla_{\mu}\nabla_{\nu}\phi-\textstyle\frac{1}{6}G_{5,X}(\phi,X) [(\Box\phi)^3\nonumber\\
	& -3\Box\phi\nabla^{\mu}\nabla^{\nu}\phi\nabla_{\mu}\nabla_{\nu}\phi   +2\nabla_{\mu}\nabla_{\nu}\phi\nabla_{\lambda}\nabla^{\mu}\phi\nabla^{\nu}\nabla^{\lambda}\phi]\label{g5}\; .
\end{align}
The coefficients $G_i(\phi,X)$ are generic differentiable functions and $X$ is the kinetic term $-\nabla^{\mu}\phi\nabla_{\mu}\phi$, in modern notation \cite{Papantonopoulos:2015cva}. As it is well known, both minimal coupling, general relativity itself, and Brans-Dicke theory are examples of particular cases of \eqref{horndact}. For recent reviews on Horndeski theories, see \cite{Kobayashi_2019,HEISENBERG20191,Amendola:2019laa}.

Another important subclass of Horndeski gravity is represented by the nonminimal derivative coupling theory, which was studied in several works of L. Amendola, S. Sushkov, and others \cite{AMENDOLA1993175,PhysRevD.80.103505,PhysRevD.81.083510,PhysRevD.85.123520,PhysRevD.88.083539}. In those papers, it is shown how the contribution from the $G_5$ family of Horndeski theories provides an inflationary mechanism in some different scenarios without a fine-tuned potential $V(\phi)$. This fact indicates that the $G_5$ family strongly affects the cosmology of the very early universe. Such term also plays an important role in the Fab Four theory \cite{PhysRevLett.108.051101,PhysRevD.85.104040,1475-7516-2012-12-026}, since its ``John'' term corresponds to \eqref{g5} for $G_5=G_5(\phi)$ after an integration by parts, and discarding surface terms, with the introduction of the notation
\begin{equation}
V_J(\phi)=\frac{d G_5(\phi)}{d\phi}\; ,
\end{equation}
which we adopt here. Another application of $G_5$ theories is for Galilean black holes \cite{Babichev2014}. Additionally, it is shown in Reference \cite{AMENDOLA1993175} that theories containing derivative couplings like the one of $G_5$ are not equivalent to an Einstein frame theory by a conformal transformation $\tilde{g}_{\mu\nu}=e^{2\omega}g_{\mu\nu}$, which implies that derivative couplings are actual alternatives to canonical scalar field cosmology. 

The specific cosmology of $G_5(\phi)$ theories, represented by Lagrangian
\begin{equation}\label{lagtot}
L=\sqrt{-g}\left[-V_J(\phi)G^{\mu\nu}\nabla_{\mu}\phi\nabla_{\nu}\phi -V(\phi)\right]\; ,
\end{equation}
was presented in a recent paper \cite{TORRES2019135003}, where \eqref{lagtot} is called the Fab Four John theory. In \eqref{lagtot}, $V$ is some potential, with both $V$ and $V_J$ generic differentiable functions of $\phi$, and $g$ is the determinant of the spatially flat Friedmann-Lema\^itre-Robertson-Walker metric:
\begin{equation}\label{metfrwn}
ds^2 = N^2dt^2 - a^2\delta_{ij}dx^idx^j\; ,
\end{equation}
where $N(t)$ is the lapse function. In \cite{TORRES2019135003}, it is shown that \eqref{lagtot} has a wide range of cosmological solutions, that may be relevant for the primordial universe, because it may be responsible, for instance, for a cosmological bounce. 

There is a debate concerning the relation between $G_5$ theories and gravitational waves, but it was already explored in several papers (see \cite{Kobayashi_2019} and \cite{Gong2018}, for instance) and, for \eqref{lagtot} in particular, it was addressed in detail in \cite{TORRES2019135003}. Taking into account that debate, one has to consider \eqref{lagtot} as a contribution to the canonical Lagrangian for the minimal coupling in order to study stages of the evolution of the universe after inflation \cite{TORRES2019135003}. But we shall not face that issue here, since we are interested in studying the universe dynamics at Planck's scales, before any inflationary period.

The quantization of \eqref{lagtot} is done in \cite{TORRES2019135003} only for very particular cases, because of the structure of its corresponding Hamiltonian:
\begin{equation}\label{ham}
H=N\left[\frac{-3p_{a}^{2/3}p_{\phi}^{2/3}}{2\sqrt[3]{6aV_J(\phi)}} +a^3V(\phi)\right]\equiv N\mathcal{H}\; .
\end{equation}
That limitation exists because of the non-integer power $2/3$. Roughly speaking, there are two simple ways to apply a quantization in such a case: canonical transformations and fractional derivatives \cite{Herrmann:2011zza}. In \cite{TORRES2019135003}, only the first option was explored, but only for very particular cases, remaining a lack for more general approaches to the quantum effects generated by \eqref{lagtot}. This happens due to the limitations of the particular canonical transformation technique applied in \cite{TORRES2019135003}, which is strongly dependent of the functional forms of $V$ and $V_J$. Indeed, for each particular form of $V$ and $V_J$, a different transformation would be necessary, thus making it impossible to have a more general overview of the quantum dynamics of the theory.

There is also a third possible solution, that we shall not consider here, which is based on the so-called bounded functional calculus \cite{reed1980methods}, that enables one to quantize suitable functions $f(p)$, where $p$ represent the momenta. With this alternative technique, it may be possible to apply a quantization to the Hamiltonian under consideration, by taking $f(p)\sim (p_{a}p_{\phi})^{2/3}$. But we shall not apply this method here because our goal in this paper is to study the simplest alternative to canonical transformations, and the bounded functional calculus would require a deeper mathematical analysis to guarantee that $f(p)$ satisfies all the required conditions for such a formalism. This may be explored in a future work, with the insight given by this first approach.

The aim of this paper is to study the second alternative: to apply a fractional calculus technique to investigate the quantum cosmology of \eqref{ham} at a deeper level. We apply the so-called conformable fractional derivative (CFD), which was proposed in \cite{KHALIL201465} to generalize Dirac quantization rule and obtain a Wheeler-DeWitt equation which represents the quantum cosmological version of \eqref{ham}, for any $V$ and $V_J$. The results thus obtained are interpreted by means of Bohm-de Broglie approach \cite{PhysRev.85.166,PhysRev.85.180}, as done in \cite{TORRES2019135003}. The motivation to apply Bohm-de Broglie comes from the argument that a more usual interpretation would be inapplicable to the universe as a whole, since there would not exist a classical exterior domain to make the measurement \cite{Omnes}, and this is considered by some authors as a limitation of standard interpretations of quantum mechanics when applied to cosmology \cite{ACACIODEBARROS1998229,PintoNeto:2004uf,10.2307/193027}. The Bohmian quantum cosmology is studied and explained in detail in \cite{PintoNeto:2004uf,0264-9381-30-14-143001}. Of course, there is no final word in the interpretation of quantum mechanics, which still is an open debate \cite{freire2014quantum}.

The paper is organized as follows. In Section \ref{sec-fab4j}, we briefly review the classical cosmology of the Fab Four John theory \cite{TORRES2019135003}, summarizing its most important aspects. In Section \ref{sec-frac-cfd}, we address the problem of the quantization with fractional powers in the momenta, pointing out some of the issues faced by the fractional derivatives, which motivates the study of the recently proposed conformable fractional derivative. In Section \ref{sec-cfd-quant}, we apply CFD to obtain a quantum cosmological version of Fab Four John, based on the deterministic Bohm-de Broglie interpretation. We then discuss what is expected from such a quantum theory, in order to establish the physical criteria to analyze its solutions. That formalism gives a quantum dynamical system, for $a(t)$ and $\phi(t)$, which can be studied in several manners. In Section \ref{sec-sol1}, we study that dynamical system for the same functional forms for $V(\phi)$ and $V_J(\phi)$ that generate the classical solutions of \cite{TORRES2019135003}, so that we can compare the classical and the quantum versions, for the same models. In Section \ref{sec-sol2}, we study the same dynamical system by another point of view: now we choose $V(\phi)$ and $V_J(\phi)$ in such a way that the scalar field can be interpreted as cosmological time and $a(t)$ has suitable forms, avoiding the fine-tuning of $V$ and $V_J$. Finally, in Section \ref{sec-conc}, we make some conclusive remarks.

\section{The Fab Four John Cosmological Theory}\label{sec-fab4j}
Let us consider \eqref{ham} in minisuperspace \cite{TORRES2019135003}. Classically, $V_J(\phi)$ and $V(\phi)$ must have the same sign for every value of $\phi$, due to the classical constraint below:
\begin{equation}
\frac{\dot{a}^2\dot{\phi}^2}{a^2}=\frac{V}{9V_J} \; .\label{eqclassn}
\end{equation}
The Hamilton equations of \eqref{ham} give the following system (see \cite{TORRES2019135003} for details):
\begin{subequations}\label{afip}
	\begin{align}
		\dot{a}=-\frac{2Na^3V}{3p_a}\; ,\label{ap}\\
		\dot{\phi}=-\frac{2Na^3V}{3p_{\phi}}\; , \label{fip}
	\end{align}
\end{subequations}
which will play a fundamental role in the Bohmian interpretation of the quantization, as we will see in Section \ref{sec-cfd-quant}.

In \cite{TORRES2019135003}, it is shown that the equations of motion obtained from the Hamilton equations of \eqref{ham} can be integrated to become the first-order system:
\begin{subequations}\label{eqelc3}
	\begin{align}
		\dot{a}&=a^{-2}V(\phi)\; ,  \label{apc}\\
		\dot{\phi}&=\textstyle\frac{1}{3}a^3[V(\phi)V_J(\phi)]^{-1/2}\; .  \label{fipc}
	\end{align}
\end{subequations}
In Table \ref{sol-plb-class-tab}, we show the functional forms of $V$ and $V_J$ studied in \cite{TORRES2019135003} and their respective solutions for $a(t)$ obtained from \eqref{eqelc3}. For all of them, we identify the scalar field as the time coordinate.

\begin{table}[H]
	\caption{Some solutions of the classical system \eqref{eqelc3}, obtained from \cite{TORRES2019135003}. The quantities $V_0,\phi_0,t_0,a_0,a_m,w,\gamma,$ and $\omega$ are positive real constants and $f(\phi)\equiv a_m+\textstyle\frac{V_0}{\omega}[1-\cos(\omega\phi)]$.}
	\label{sol-plb-class-tab}
	\centering
	\begin{tabular}{cccc}
		\toprule
		& $\mathbf{V_J}$	& $\mathbf{V}$ & $\mathbf{a(t)}$\\
		\midrule
		(i)	& $\textstyle\frac{V_0}{4}(1+w)^2\phi^{\frac{3+w}{1+w}}$ & $V_0\phi^{\frac{1-w}{1+w}}$ & $(t/t_0)^{\frac{2}{3(1+w)}}$ \\ 
		(ii)	& $\frac{V_0}{9\gamma^2}e^{3\gamma\phi}$ & $V_0e^{3\gamma\phi}$ & $a_0e^{\gamma t}$ \\ 
		(iii)	& $\frac{V_0}{4\phi}(1+w)^2(\phi_{0}^{2}+\phi^2)^{\frac{2+w}{1+w}}$ & $V_0\phi(\phi_{0}^{2}+\phi^2)^{\frac{-w}{1+w}}$ & $a_0[1+(t/\phi_0)^2]^{\frac{1}{3(1+w)}}$ \\ 
		(iv)	& $\frac{V_0\cosh^4(\gamma\phi)}{9\gamma^2 \sinh(\gamma \phi)}$ & $V_0\sinh(\gamma\phi)\cosh^2(\gamma\phi)$ & $a_0\cosh(\gamma t)$ \\ 
		(v)	& $\frac{f^4(\phi)}{9V_0\sin(\omega\phi)}$ & $V_0\sin(\omega\phi)f^2(\phi)$ & $a(t)=a_m+\frac{V_0}{\omega}[1-\cos(\omega t)]$ \\ 
		\bottomrule
	\end{tabular}
\end{table}

In Table \ref{sol-plb-class-tab}, the solution (i) represents the analogous of the perfect fluid cosmology, and thus $w$ is the analogous of the equation of state parameter: for $w=0$, the universe is dominated by dust; for $w=1$, the universe is dominated by stiff matter; if $w=1/3$, the universe is dominated by radiation. The solution (ii) represents a de Sitter universe, with accelerated expansion. The solution (i) is singular at $t=0$ and the solution (ii) may be considered or not as a singular solution, as discussed, for instance, in \cite{Wetterich:2019vws,Anabalon2019}. The solutions (iii) and (iv) are non-singular bouncing solutions that, for $t\gg0$, approach (i) and (ii), respectively. Finally, (v) represents a cyclic universe. 


The above description accounts for the classical cosmology of the theory. However, when we consider the very early universe, quantum effects may be taken into account for a more complete description \cite{Calcagni:2017sdq,bojowald2011quantum}. But in the present case of Hamiltonian \eqref{ham}, quantization itself constitutes an issue, because of the term $(p_ap_{\phi})^{2/3}$. Thanks to the fractional power in the momenta, the quantization rule cannot be directly and unambiguously applied, since the momentum would become a derivative of fractional order. 

There exist two main ways to deal with that issue: with a {\it canonical transformation} to obtain a new Hamiltonian, for which there are only integer powers in the (new) momenta; or with a {\it fractional derivative} to generalize the quantization rule for momenta with non-integer powers \cite{Herrmann:2011zza}. In that sense, canonical transformations have two advantages: first, it may happen that the new Hamiltonian has no fractional powers in the momenta at all, in which case the fractional powers should be understood as a variable-dependent problem, not as a real structural feature of the theory; and second, they automatically guarantee that the equations of motion for both the old and the new variables are equivalent.

Although, for the Fab Four John theory, the canonical transformations have also the disadvantage of being strongly dependent of the functional forms of $V$ and $V_J$, so that only two classical solutions (namely, (i) and (ii), from Table \ref{sol-plb-class-tab}) were studied at the quantum level by means of a canonical transformation \cite{TORRES2019135003}. This, of course, does not mean that the canonical transformations themselves are problematic, since it is well known that they are a fundamental well-studied formalism from classical mechanics. But, from the point of view of the quantization of Fab Four John, that technique seems to restrict the range of models (each choice of $V$ and $V_J$ gives a different model) that can be studied in the same footing. Another characteristic of the classical solutions of \cite{TORRES2019135003} that justifies a quantization approach is: their classical bounce solutions should have a more fundamental justification. This is a consequence of the fact that if there was a bounce, then it should have happened when the universe was at Planck's scales \cite{NOVELLO2008127}, precisely when the classical theories begin to need corrections. 

Therefore, one way to have a wider comprehension of Fab Four John theory at a quantum level is to extend quantization through some fractional derivative, that can be applied, in principle, for any smooth $V$ and $V_J$. With such a formalism, the Dirac quantization rule,
\begin{equation}\label{quantcan}
p_{j}^{n}\longrightarrow\hat{p}_{j}^{n}=(-i\hbar)^{n}\partial_{j}^{n}\; ,
\end{equation}
makes sense even when $n$ is not an integer, and thus the term $p_{a}^{2/3}p_{\phi}^{2/3}$ becomes a fractional derivative operator. Hence, we come up with a fractional Wheeler-DeWitt equation $\hat{\mathcal{H}}\psi=0$. Even though this procedure is ambiguous, because there are several definitions of fractional derivatives, the classical formalism reviewed above together with Bohmian interpretation of its quantum cosmology provides a mechanism to address the physical viability of a given fractional derivative. This will be clarified in Section \ref{sec-cfd-quant}.

\section{Fractional Derivatives and CFD}\label{sec-frac-cfd}
The higher order derivatives from usual calculus have integer orders and the fractional derivatives are generalizations of those for non-integer orders. There are, in fact, several different definitions for them, and the most popular ones are those defined by integrals, like Riemann-Liouville, defined as follows \cite{Herrmann:2011zza,Malinowska:2015:AMF:2800499}: the Riemann-Liouville derivative of order $r\in[n-1,n)$ (where $n$ is a positive integer) of the function $f$ at $x$ is:
\begin{equation}\label{rl-deriv}
D_{a}^{r}(f)(x)=\frac{1}{\Gamma(n-r)}\frac{d^n}{dx^n}\int_{a}^{x}\frac{f(\xi)\,d\xi}{(x-\xi)^{r-n+1}}\; .
\end{equation}
The fractional calculus has several applications in physics. For example, they are applied in anomalous diffusion dynamics \cite{METZLER-2000}, wave propagation in viscoelastic media \cite{mainardi2010fractional}, lossy partial differential acoustic wave equations \cite{holm}, fractional quantum mechanics \cite{LASKIN2000298,PhysRevE.62.3135,PhysRevE.66.056108}, and to write a generalization of the conservation of mass to represent non-linear flux in a control volume \cite{WHEATCRAFT20081377}.

Even though the range of applications of fractional derivatives is wide, there are several different definitions, and there is no one that can be considered as the ``ultimate'' definition, since if a fractional derivative has some desirable properties, it certainly will not have some other important features. Thus, in order to apply fractional calculus to a physics problem, it is necessary to investigate what should be the most adequate derivative to the problem, reducing arbitrariness as much as possible.

The fractional derivatives considered above and in the cited references are defined by integrals. As a consequence, they hardly respect some basic rules of the ordinary derivatives: the fractional derivative of a constant may not be zero and the Leibniz rule is not valid in general. Therefore, in \cite{KHALIL201465} the conformable fractional derivative has been proposed, in order to cover those rules and simplify fractional calculus. For a real function $\psi$ of one real variable, the CFD of order $r$, where $0<r\leq1$, is defined by:
\begin{equation}\label{def-cfd}
\frac{d^r\psi}{dx^r}=\lim_{h\rightarrow0}\frac{\psi(x+h\cdot x^{1-r})-\psi(x)}{h}=x^{1-r}\frac{d\psi}{dx}\; .
\end{equation}
The first equation in \eqref{def-cfd} is the definition of the CFD and the second one is a property that follows immediately from the definition, and is actually equivalent to it. It is straightforward to show that the CFD is linear, satisfies the Leibniz rule, and the derivative of any order $0<r\leq1$ of a constant is zero. That derivative can be generalized for real functions $\psi$ of several variables \cite{JOUR,ALMEIDA-OPEN-MATH-2016}:
\begin{equation}\label{dfcaexx}
\frac{\partial^r\psi}{\partial x_{j}^{r}}=\lim_{h\rightarrow0} \frac{\psi(x+h\cdot x_{j}^{1-r}e_j)-\psi(x)}{h} = x_{j}^{1-r}\frac{\partial\psi}{\partial x_j}\; ,
\end{equation}
where $x=\sum x_je_j$ and $\{e_1,\ldots , e_n\}$ is the canonical basis of euclidean space $\mathbb{R}^n$. The next step is to generalize it for complex functions, but this is actually trivial: we can consider $\psi$ as complex in the limits above, or we can simply define, for a complex $\psi$ of several variables,
\begin{equation}
\frac{\partial^r\psi}{\partial x_{j}^{r}}=\frac{\partial^r}{\partial x_{j}^{r}}{\rm Re}\,\psi + i \frac{\partial^r}{\partial x_{j}^{r}}{\rm Im}\,\psi\; ,
\end{equation}
which is equivalent. Here, ${\rm Re}\,\psi$ and ${\rm Im}\,\psi$ denote the real and imaginary parts of $\psi$, respectively. For a general $r$, the domain must be restricted to a region in which all variables are positive \cite{KHALIL201465}. But, in the case of interest here $r=2/3$, so that the variables $a$ and $\phi$ can, in principle, assume any real value (with $a>0$, by the physical meaning of $a$). Also note that it follows from \eqref{dfcaexx} that a fractional differential equation reduces to a usual partial (or ordinary, if $n=1$) differential equation by the application of CFD. The generalization for $r>1$ is also in \cite{KHALIL201465}, but it will not be necessary here.

Potential applications were discussed in \cite{ZhaoLuo2017}. In \cite{ChungWon2015}, the fractional classical mechanics was constructed using CFD. In \cite{ESLAMI2016141}, that derivative was applied to solve a system of fractional coupled nonlinear Schr\"odinger equations. It has also been applied in Optics \cite{EKICI201610879}. In line with those applications, and because of the properties shown above, in this paper we will apply CFD in the quantization of Fab Four John as a mathematical tool to investigate the features of the quantum dynamics of that theory. As a last comment on applications, we would like to emphasize that it is not the goal of this paper to give a final answer for what is the ``right'' fractional derivative, or to advocate some particular definition. Instead, we are showing how the quantization of Fab Four John provides a method to study a fractional derivative and to investigate its physical implications. Therefore, this technique may be applied to other fractional derivatives in the future.

\section{Quantization of Fab Four John theory with CFD}\label{sec-cfd-quant}
We can now generalize Dirac quantization rule \eqref{quantcan} to non-integer orders using CFD. It follows from \eqref{quantcan} and \eqref{dfcaexx} that we can write
\begin{equation}\label{p23p23}
\hat{p}_{a}^{2/3}\hat{p}_{\phi}^{2/3}\psi=\hbar^{4/3}(a\phi)^{1/3}\frac{\partial^{2}\psi}{\partial a\partial\phi}\; .
\end{equation}
Hence, the Wheeler-DeWitt equation $\hat{\mathcal{H}}\psi=0$ for CFD must be, in the trivial ordering:
\begin{equation}\label{eqwdwvfdf}
\hbar^{4/3}\frac{\partial^{2}\psi}{\partial a\partial\phi}=a^3u(\phi)\psi,\qquad\mbox{where}\qquad u(\phi)\equiv V(\phi)\left[\frac{16V_J(\phi)}{9\phi}\right]^{1/3}\; .
\end{equation}
The solution $\psi(a,\phi)$ is known in quantum cosmology as the wave function of the Universe, because it determines the dynamics of both the scale factor and the scalar field.

We can easily see that the stationary classical Hamilton-Jacobi equation corresponding to Hamiltonian \eqref{ham} is
\begin{equation}\label{hjclass}
-\frac{3/2a^3}{[6aV_J(\phi)]^{1/3}}\bigg( \frac{\partial S}{\partial a}\cdot\frac{\partial S}{\partial\phi} \bigg)^{2/3}+V(\phi)=0\; ,
\end{equation}
where $S$ is the Hamilton principal function. Now, we will show that this equation has a quantum contribution coming from the quantization with CFD applying the Bohm method. In a more canonical case like basic quantum mechanics \cite{holland_1993}, to apply Bohm-de Broglie interpretation we have to write $\psi=R(a,\phi)e^{iS(a,\phi)/\hbar}$, where $R$ and $S$ are real functions without explicit $\hbar$. That procedure transforms, for instance, the Schr\"odinger equation for a single particle
\begin{equation}\label{schr}
-\frac{\hbar^2}{2m}\nabla^2\psi+V(\mathbf{x})\psi=i\hbar\frac{\partial\psi}{\partial t}\;,
\end{equation}
into two equations
\begin{subequations}
	\label{eqscr}
	\begin{align}
		\frac{\partial R^2}{\partial t}+\nabla\cdot\bigg(\frac{R^2\nabla S}{m}\bigg) &= 0\;, \label{schrim}\\
		\frac{\partial S}{\partial t} +\frac{|\nabla S|^2}{2m}+V(\mathbf{x})+Q(\mathbf{x})&=0\;, \label{schrre}
	\end{align}
\end{subequations}
and the fact that $R$ and $S$ do not have an explicit $\hbar$ dependence is what ensures that the term
\begin{equation}\label{potqschr}
Q(\mathbf{x})=-\frac{\hbar^2}{2m}\frac{\nabla^2R}{R}\; .
\end{equation}
is of order $\hbar^2$. Thus, we can say that any effects of $Q$ are relevant only for an energy scale for which $\hbar^2$ is relevant, no matter what particular wave function $\psi$ (solution of \eqref{eqwdwvfdf}) is considered.  The quantity $Q$ is called then the quantum potential. But that $\hbar$-dependence is not necessarily respected, even for some solutions of \eqref{schr} (see \cite{holland_1993}). Thus, the $\hbar$-dependence is important as a rough idea of the overall dynamics, but it should not be taken as literally determinant.

A simple way to adapt the polar expansion $\psi=Re^{iS/\hbar}$ to \eqref{eqwdwvfdf} is to write
\begin{equation}\label{exp-frac-wdw}
\psi=R\;\exp\bigg(\frac{iS}{\hbar^{2/3}}\bigg)\; ,
\end{equation}
where $R$ and $S$ are real functions. In fact, applying \eqref{exp-frac-wdw} to \eqref{eqwdwvfdf}, its real part can be writen as
\begin{equation}\label{hjquant}
-\frac{3/2a^3}{[6aV_J(\phi)]^{1/3}}\bigg( \frac{\partial S}{\partial a}\cdot\frac{\partial S}{\partial\phi} \bigg)^{2/3}+V(\phi)+Q(a,\phi)=0\; ,
\end{equation}
where
\begin{equation}\label{qp-geral-fj4}
Q(a,\phi)=-V(\phi)+\frac{3/2a^3}{[6aV_J(\phi)]^{1/3}}\bigg[ \frac{\hbar^{4/3}}{R}\frac{\partial^2R}{\partial a\partial\phi}-a^3u(\phi) \bigg]^{2/3}\; .
\end{equation}
In analogy with the Bohmian interpretation for the simplest case of a single particle described by the Schr\"odinger equation, we can compare equations \eqref{hjclass} and \eqref{hjquant} to conclude that $Q$ can be understood as the quantum potential of the system, since it is the effective contribution to the classical Hamilton-Jacobi equation \eqref{hjclass}. The presence of the classical potential $V$ in the expression \eqref{qp-geral-fj4} for the quantum potential is not a problem, since this happen, for instance, for the simple case of quantum harmonic oscillator in Schr\"odinger equation \cite{holland_1993}, and no contradiction with the numerical results from standard interpretation is created. Hence, we can also conclude that we can identify the quantum phase $S$ with the Hamilton principal function. In other words, the relations
\begin{equation}\label{eqguiaafi}
p_a=\frac{\partial S}{\partial a}\qquad\mbox{and}\qquad p_{\phi}=\frac{\partial S}{\partial\phi}
\end{equation}
still are valid for the quantization with CFD. Thus, it follows from \eqref{afip} and \eqref{eqguiaafi} that the Bohmian guidance equations are
\begin{equation}\label{statqds}
\dot{a}=-\frac{2}{3}Na^3V\bigg( \frac{\partial S}{\partial a} \bigg)^{-1} \qquad\mbox{and}\qquad \dot{\phi}=-\frac{2}{3}Na^3V \bigg( \frac{\partial S}{\partial\phi} \bigg)^{-1}\; .
\end{equation}
Now we have to solve Wheeler-DeWitt equation \eqref{eqwdwvfdf} to obtain a particular $S$, and the quantum dynamics of $a(t)$ and $\phi(t)$ will be completely determined by the autonomous dynamical system \eqref{statqds} thus obtained.

The Wheeler-DeWitt equation \eqref{eqwdwvfdf} can be solved for any $V$ and $V_J$, by the method of separation of variables, and has the basic solution
\begin{equation}\label{psik}
\psi(a,\phi)=\exp\bigg[\frac{i}{\hbar^{2/3}}\bigg(-\frac{k}{4}a^{4}+\frac{1}{k}\int^{\phi} u(\bar{\phi})d\bar{\phi}\bigg)\bigg]\; ,
\end{equation}
where $k\in\mathbb{R}-\{0\}$ is the separation constant and $\int^{\phi} u(\bar{\phi})d\bar{\phi}$ denotes the primitive of $u(\phi)$ for which the integration constant is zero. For solution \eqref{psik}, $R=1$ and $S=-\frac{k}{4}a^4+\frac{1}{k}\int u$, from which follows that the quantum potential is
\begin{equation}\label{qp-sw}
Q(a,\phi)=-V(\phi)+\bigg[\frac{9V^6(\phi)}{16a^{12}V_J(\phi)\phi^2}\bigg]^{1/9}\; ,
\end{equation}
and the guidance equations are
\begin{subequations}\label{guidk}
	\begin{align}
		\dot{a}&=\frac{2}{3k}V(\phi)\;,\label{guidk-a}\\
		\dot{\phi}&= -\frac{2ka^3}{3}\bigg[\frac{9\phi}{16V_J(\phi)} \bigg]^{1/3}\; .\label{guidk-phi}
	\end{align}
\end{subequations}
Those quantum guidance equations form an autonomous dynamical system, very similar in form to the classical one \eqref{eqelc3}. The difference between them is, of course, a direct consequence of the quantization. To make that clear, we must set $V$ and $V_J$ and then evaluate the corresponding quantum potential, which is given by \eqref{qp-sw}. With the deterministic $Q(t)$ and $a(t)$ thus obtained, we can investigate their physical meaning from the point of view of Bohmian interpretation. This is done observing that the time intervals when $Q$ dominates are the periods when the quantum effects generated by the quantization method presented here are more relevant. There is indeed an enormous freedom in \eqref{guidk}, so we will restrict the discussion to some relevant examples.

In the two next sections, we study \eqref{guidk} for the scenarios of Table \ref{sol-plb-class-tab}, in two ways. First, to explore the intrinsic differences between  the classical dynamics \eqref{eqelc3} and the quantum one \eqref{guidk}, we study each combination of $V$ and $V_J$ from Table \ref{sol-plb-class-tab}, in Section \ref{sec-sol1}. In that first way, we will see that the domination of $Q$ is very strong for all the scenarios, and hence the solutions are really different from the classical ones. This motivates us to study \eqref{guidk} in an alternative way, explored in Section \ref{sec-sol2}: to set $V$ and $V_J$ in such a way that each one of the solutions for $a(t)$ itself in Table \ref{sol-plb-class-tab} are obtained. For simplicity, we will consider in Section \ref{sec-sol2} that the scalar field is time itself, in analogy with what was done for the classical system \eqref{eqelc3}, in \cite{TORRES2019135003}. As we know, in order to consider the scalar field as a time scale, we must ensure that it is a monotonic function of time. This is true if, and only if, the derivative $\dot{\phi}$ is always positive or always negative. By \eqref{guidk-phi}, this is equivalent to the requirement that either both $\phi$ and $V_J$ have always the same sign or else they have always opposite signs. Thus, this imposes a restriction on the sign of $V_J$. But that restriction is not problematic, because if, for instance, the simplest condition $\phi=t$ is imposed, that condition will be naturally satisfied. This can be checked looking at equations \eqref{vj-61}, \eqref{vj-62}, \eqref{vj-63}, \eqref{vj-64}, and \eqref{vj-65}.

Those two different ways to study the same system will provide us a catalog of solutions to shade some light over the new method of quantization based on Bohmian mechanics and fractional calculus. 

\section{Quantum Solutions I}\label{sec-sol1}
In order to compare the formalism here developed with the classical one of \cite{TORRES2019135003}, we must analyze the dynamical systems obtained from \eqref{guidk} with the functions $V$ and $V_J$ corresponding to each solution of Table \ref{sol-plb-class-tab}.

\subsection{Solution (i)}\label{subsec-1-i}
The functions $V$ and $V_J$ for the (i) solution from Table \ref{sol-plb-class-tab} give the following guidance equations:
\begin{subequations}\label{eqguias-sol-i-i}
	\begin{align}
		\dot{a} &=\frac{2V_0}{3k}\phi^{\frac{1-w}{1+w}}\; ,\\
		\dot{\phi} &=-\frac{2ka^3}{3}\bigg[ \frac{9}{4V_0(1+w)^2} \bigg]^{1/3}\phi^{\frac{-2}{3(1+w)}}\;.
	\end{align}
\end{subequations}
Thus, the quantum potential, given by \eqref{qp-sw}, is:
\begin{equation}\label{qp-sol-i-i}
Q(a,\phi)=-V_0\phi^{\frac{1-w}{1+w}}+\bigg[ \frac{9V_{0}^{5}}{4(1+w)^2} \bigg]^{1/9}\phi^{\frac{1-9w}{9(1+w)}}a^{-4/3}\;.
\end{equation}

\begin{figure}[H]
	\centering
	\includegraphics[width=14.5cm]{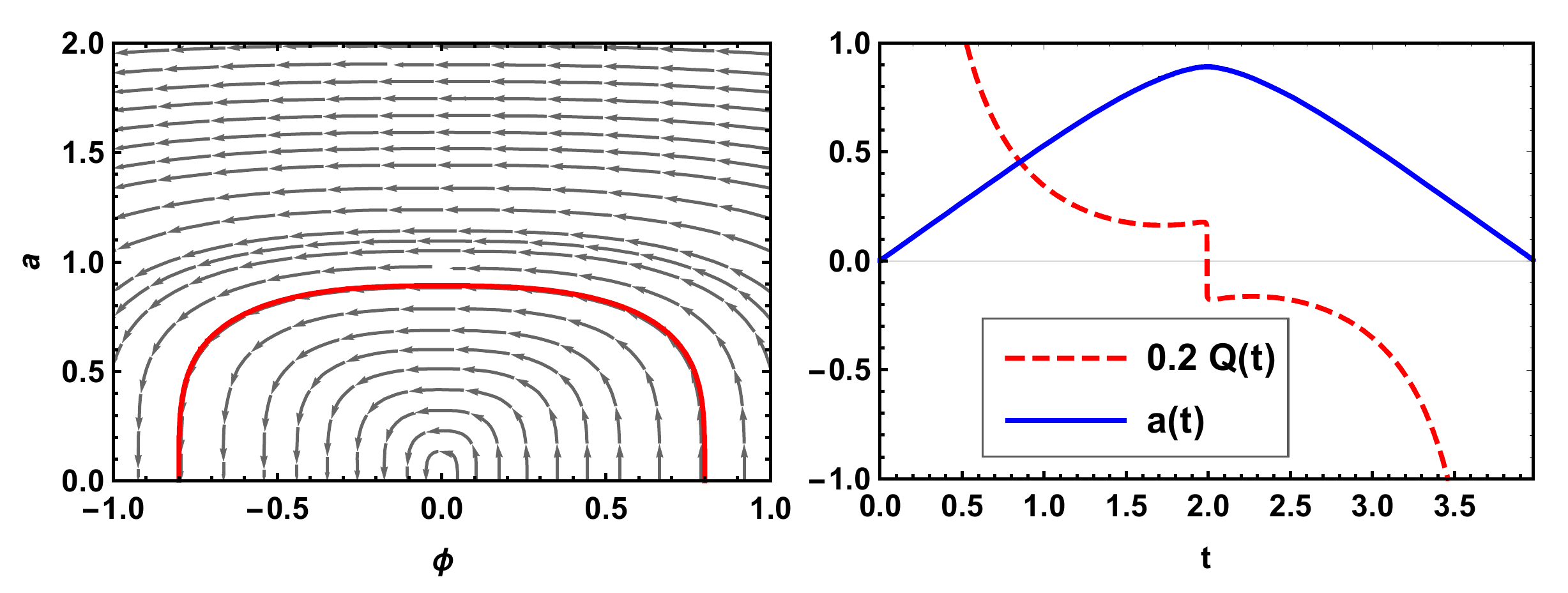}
	\caption{On the left-hand side, we see the phase portrait of system \eqref{eqguias-sol-i-i}, for the dust case ($w=0$), with $V_0=k=1$, and the red line represents the particular solution for which $\phi(0)=0.8$ and $a(0)=0.001$, as an example. On the right-hand side, we see the behavior of the scale factor (thick line) versus the quantum potential \eqref{qp-sol-i-i} (dashed line), both being evaluated for the particular solution highlighted in the phase portrait.}
	\label{sol-i-i-label}
\end{figure}

The equations \eqref{eqguias-sol-i-i} form a dynamical system. For $w=0$, its phase portrait is sketched in Figure \ref{sol-i-i-label}, where we can see that the solutions are all singular: the universe starts with an expansion from the singularity $a=0$, then, at some point, it reaches a maximum, and finally it evolves back to the singularity. Also in Figure 
\ref{sol-i-i-label}, we can see the evolution of the quantum potential for a particular solution, indicating that $Q$ dominates when $a$ is near the singularity $a=0$. Comparing now the quantum solutions with the classical ones, we can conclude that the functions $V$ and $V_J$ that provide the (i) classical expanding solution $a\sim t^{2/3}$ from Table \ref{sol-plb-class-tab} do not describe an expanding universe when applied to the fractional quantum formalism. Therefore, the system is singular and the quantum potential dominates around the singularity $a=0$, which means that the quantum effect in this particular case is to make the universe go back to singularity. All those comments are still valid for the cases $w=1$ and $w=1/3$.

\subsection{Solution (ii)}\label{subsec-1-ii}
Now, for the $V$ and $V_J$ of the solution (ii) from Table \ref{sol-plb-class-tab}, the guidance equations are:
\begin{subequations}\label{eqguias-sol-i-ii}
	\begin{align}
		\dot{a} &=\frac{2V_0}{3k}e^{3\gamma\phi}\; ,\\
		\dot{\phi} &=-k\bigg[ \frac{3\gamma^2\phi}{2V_0} \bigg]^{1/3}a^3e^{-\gamma\phi}\;.
	\end{align}
\end{subequations}
And the quantum potential is:
\begin{equation}\label{qp-sol-i-ii}
Q(a,\phi)=-V_0e^{3\gamma\phi}+\bigg( \frac{81\gamma^2V_{0}^{5}}{16\phi^2a^{12}} \bigg)^{1/9}e^{5\gamma\phi/3}\;.
\end{equation}

\begin{figure}[H]
	\centering
	\includegraphics[width=14.5cm]{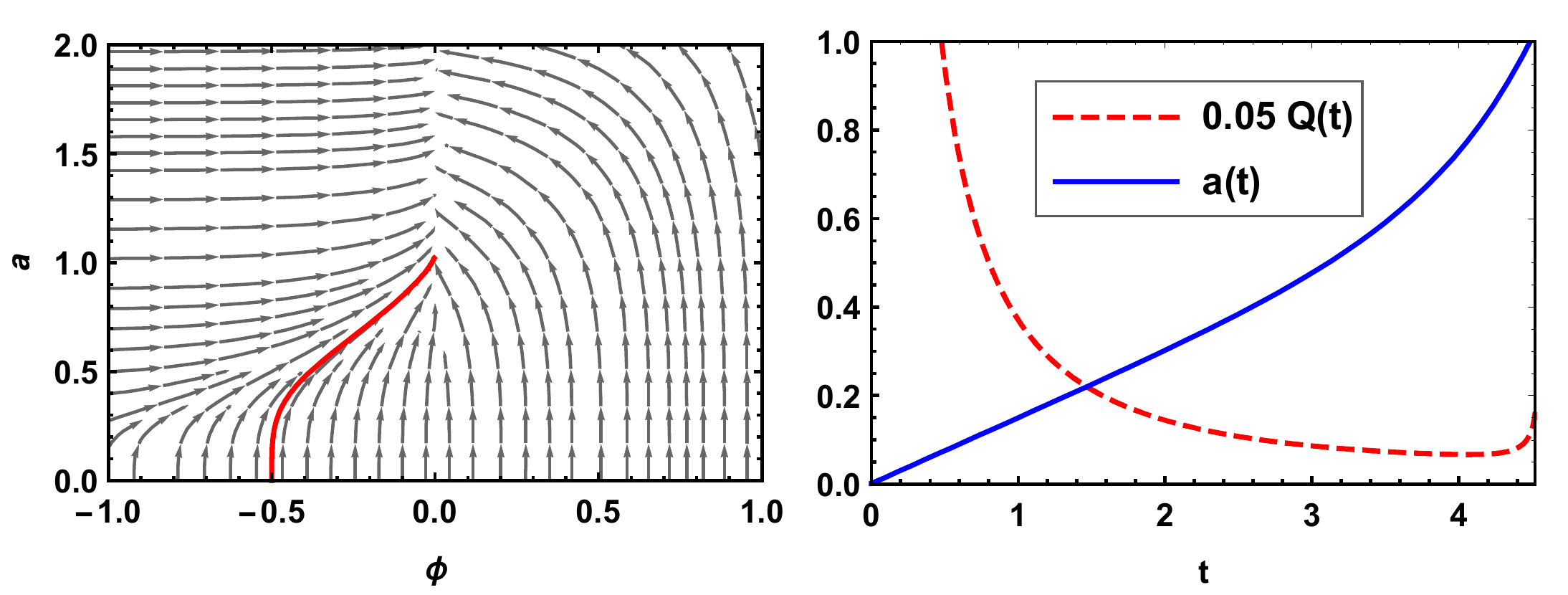}
	\caption{On the left-hand side, we see the phase portrait of system \eqref{eqguias-sol-i-ii}, with $V_0=k=\gamma=1$, and the red line represents the  solution for which $\phi(0)=-0.5$ and $a(0)=0.001$, as an example. On the right-hand side, we see the behavior of the scale factor (thick line) and the quantum potential \eqref{qp-sol-i-ii} (dashed line), both being evaluated for the particular solution highlighted in the phase portrait.}
	\label{sol-i-ii-label}
\end{figure}

In Figure \ref{sol-i-ii-label}, we can see the phase portrait of dynamical system \eqref{eqguias-sol-i-ii} and its comparison with the quantum potential, for a particular solution. It can be seen that all the solutions are singular. Although they represent expanding universes, all of them evolve to $\phi=0$, which is a singularity of \eqref{eqguias-sol-i-ii}. Thus, the solutions are very different from the classical ones $a\sim e^{\gamma t}$, which are obtained from the same pair $V$, $V_J$ (see Table \ref{sol-plb-class-tab}). The behavior of $Q$ is very similar to the previous case.

\subsection{Solution (iii)}\label{subsec-1-iii}
For the $V$ and $V_J$ of solution (iii) from Table \ref{sol-plb-class-tab}, the quantum dynamical system is the following:
\begin{subequations}\label{eqguias-sol-i-iii}
	\begin{align}
		\dot{a} &=\frac{2V_0}{3k}\phi(\phi_{0}^{2}+\phi^{2})^{\frac{-w}{1+w}}\; ,\\
		\dot{\phi} &=-\frac{2ka^3}{3}\bigg[ \frac{9\phi^2}{4V_0(1+w)^2} \bigg]^{1/3} (\phi_{0}^{2}+\phi^{2})^{-\frac{2+w}{3(1+w)}}	\;.
	\end{align}
\end{subequations}
Thus, the quantum potential is given by:
\begin{equation}\label{qp-sol-i-iii}
Q(a,\phi)=-V_0\phi(\phi_{0}^{2}+\phi^{2})^{\frac{-w}{1+w}}+\bigg[\frac{9V_{0}^{5}\phi^5}{4(1+w)^2a^{12}}\bigg]^{1/9} (\phi_{0}^{2}+\phi^{2})^{-\frac{2+7w}{9(1+w)}}\;.
\end{equation}

\begin{figure}[H]
	\centering
	\includegraphics[width=14.5cm]{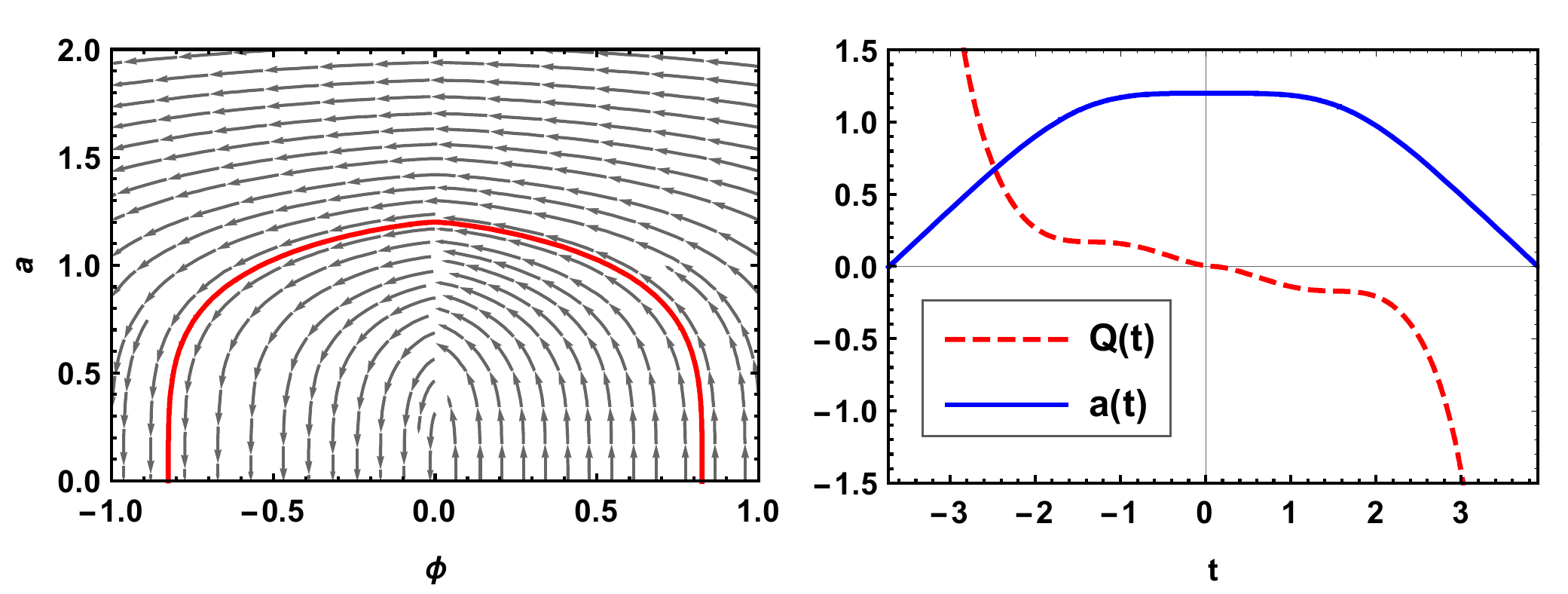}
	\caption{The phase portrait of system \eqref{eqguias-sol-i-iii}, for the dust case ($w=0$) is shown on the left-hand side, with $V_0=\phi_0=k=1$, and the red line represents the  solution for which $\phi(0)=0.0001$ and $a(0)=1.2$, as an example. On the right-hand side, we see the behavior of the scale factor (thick line) and the quantum potential \eqref{qp-sol-i-iii} (dashed line), both being evaluated for the particular solution highlighted in the phase portrait.}
	\label{sol-i-iii-label}
\end{figure}

For the classical case, shown in the solution (iii) of Table \ref{sol-plb-class-tab}, the scale factor bounces and then goes to an expansion. From Figure \ref{sol-i-iii-label}, we can see that this is not the case for the quantum solutions of \eqref{eqguias-sol-i-iii} for $w=0$ because they are all singular: the solutions start from the singularity $a=0$, reach a maximum at $\phi=t=0$ and then contract back to the singularity $a=0$. The quantum potential above can be interpreted in the same way as \eqref{qp-sol-i-i} was. All those comments are still valid for the cases $w=1$ and $w=1/3$.

\subsection{Solution (iv)}\label{subsec-1-iv}
The $V$ and $V_J$ of solution (iv) from Table \ref{sol-plb-class-tab} give:
\begin{subequations}\label{eqguias-sol-i-iv}
	\begin{align}
		\dot{a} &=\frac{2V_0}{3k}\sinh(\gamma\phi)\;\cosh^2(\gamma\phi)\; ,\\
		\dot{\phi} &=-ka^3\bigg[ \frac{3\gamma^2\phi\sinh(\gamma\phi)}{2V_0\cosh^4(\gamma\phi)} \bigg]^{1/3}	\;.
	\end{align}
\end{subequations}
Hence, the quantum potential is given by:
\begin{equation}\label{qp-sol-i-iv}
Q(a,\phi)=-V_0\sinh(\gamma\phi)\cosh^2(\gamma\phi)+\bigg[ \frac{81\gamma^2V_{0}^{5}\sinh^7(\gamma\phi)\cosh^8(\gamma\phi)}{16\phi^2a^{12}} \bigg]^{1/9}\;.
\end{equation}

\begin{figure}[H]
	\centering
	\includegraphics[width=14.5cm]{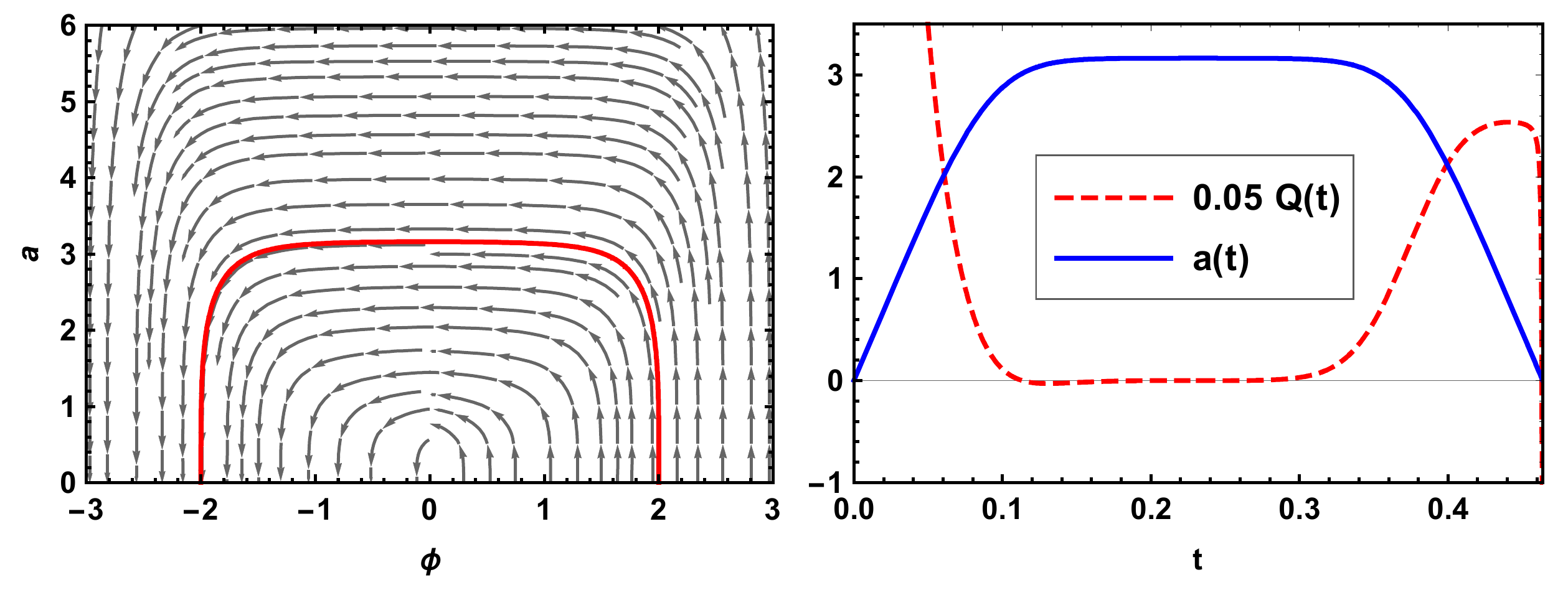}
	\caption{The phase portrait of system \eqref{eqguias-sol-i-iv} is shown on the left-hand side, with $V_0=\gamma=k=1$, and the red line represents the  solution for which $\phi(0)=2$ and $a(0)=0.001$, as an example. On the right-hand side, we see the behavior of the scale factor (thick line) and the quantum potential \eqref{qp-sol-i-iv} (dashed line), both being evaluated for the particular solution highlighted in the phase portrait.}
	\label{sol-i-iv-label}
\end{figure}

We can see the phase portrait of\eqref{eqguias-sol-i-iv} in Figure \ref{sol-i-iv-label}, where it is evident that all solutions are singular. We can also see in Figure \ref{sol-i-iv-label} the quantum potential behaves like the previous cases: it diverges on the singularity $a=0$.

\subsection{Solution (v)}\label{subsec-1-v}
Finally, for the $V$ and $V_J$ of solution (v) from Table \ref{sol-plb-class-tab}, we find the following dynamical system:
\begin{subequations}\label{eqguias-sol-i-v}
	\begin{align}
		\dot{a} &=\frac{2V_0}{3k}\sin(\omega\phi)f^2(\phi)\; ,\\
		\dot{\phi} &=-ka^3\bigg[ \frac{3V_0\phi\sin(\omega\phi)}{2f^4(\phi)} \bigg]^{1/3}\;,
	\end{align}
\end{subequations}
where $f(\phi)=a_m+\textstyle\frac{V_0}{\omega}[1-\cos(\omega\phi)]$. The quantum potential is given by:
\begin{equation}\label{qp-sol-i-v}
Q(a,\phi)=-V_0\sin(\omega\phi)f^2(\phi)+\bigg[ \frac{81V_{0}^{7}\sin^7(\omega\phi)f^8(\phi)}{16a^{12}\phi^2} \bigg]^{1/9}\;.
\end{equation}

\begin{figure}[H]
	\centering
	\includegraphics[width=14.5cm]{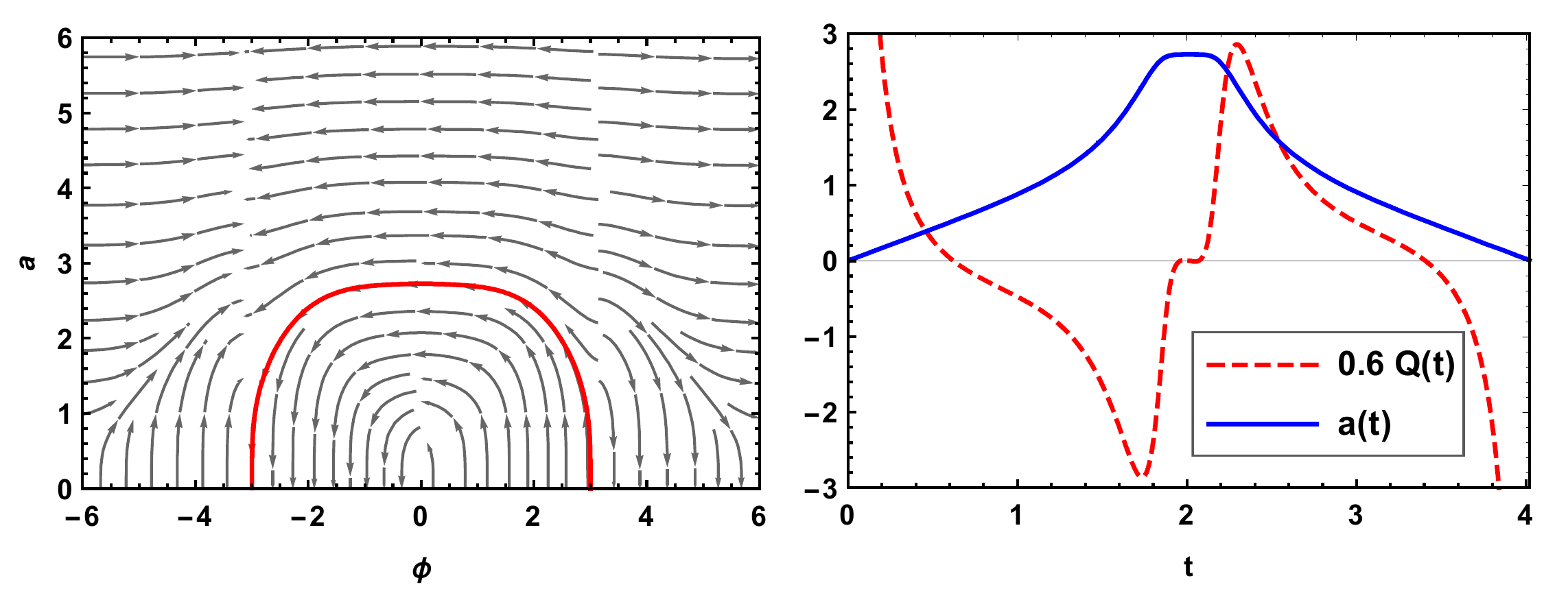}
	\caption{The phase portrait of system \eqref{eqguias-sol-i-v} is shown on the left-hand side, with $V_0=\omega=k=1$, and the red line represents the  solution for which $\phi(0)=3$ and $a(0)=0.001$, as an example. On the right-hand side, we see the behavior of the scale factor (thick line) and the quantum potential \eqref{qp-sol-i-v} (dashed line), both being evaluated for the particular solution highlighted in the phase portrait.}
	\label{sol-i-v-label}
\end{figure}

In Figure \ref{sol-i-v-label}, we see the phase portrait of the guidance equations \eqref{eqguias-sol-i-v} and the behavior of the quantum potential for a particular solution. Again, all solutions are singular and the quantum potential diverges when $a=0$.

From all the solutions studied in this section, we can say that the same choices for $V$ and $V_J$ made in the classical theory are not able to give non-singular solutions. This is a consequence of the intrinsic differences between the classical system \eqref{eqelc3} and the quantum one \eqref{guidk}, which is confirmed by the behavior of $a(t)$ and $Q(t)$ evaluated above, since they all exhibit very strong quantum effects. In other words, the quantum dynamics is not trivial, in the sense that it not just a reproduction of the classical dynamics. Although, the five solutions above do not describe expanding universes, and they are all singular. Thus, we need to go further and investigate a way to obtain non-singular solutions from  \eqref{guidk} that may provide not just strong quantum effects but, instead, quantum effects that avoid the singularity, with a reasonable physical justification for that, through the quantum potential.

\section{Quantum Solutions II}\label{sec-sol2}
Now we analyze the problem by another point of view: since equations \eqref{eqelc3} and \eqref{guidk} are different, we can ask what should be the particular expressions for $V$ and $V_J$ that would lead to the solutions for $a(t)$ in the forth column of Table \ref{sol-plb-class-tab} when the quantum system \eqref{guidk}  governs the dynamics, instead of \eqref{eqelc3}.

One could argue that that approach leads to another fine-tuning problem, very similar to the inflation one we were trying to avoid. But this is really not the case, and that can be concluded by taking a closer look to \eqref{guidk}: considering again a general $\phi$, and supposing that a desirable $a(t)$ is given, equation \eqref{guidk-a} does not determine $V(\phi)$, but only $V(\phi(t))$, which is actually a very different information, since it will depend on $\phi(t)$. But $\phi(t)$ will depend of both $a$ and $V_J$, in view of \eqref{guidk-phi}, and $V_J$ stills free. Therefore, there is no fine-tuning over $V$ and $V_J$  and the only reason to take $\phi=t$ is, as we said above, for simplicity.

\subsection{Solution (i)}\label{subsec61}
If we choose
\begin{subequations}\label{vvj-s2i}
	\begin{align}
		V_J(\phi)&=-V_{j0}\phi^{\frac{7+w}{1+w}}		\; ,\label{vj-61}\\
		V(\phi)&=V_0\phi^{-\frac{1+3w}{3(1+w)}}		\; ,
	\end{align}
\end{subequations}
then we obtain the scale factor analogous to the one of a perfect fluid cosmology: $a(t)=(t/t_0)^{\frac{2}{3(1+w)}}$, for $V_{j0}=\textstyle\frac{1}{6}k^3t_{0}^{\frac{-6}{1+w}}$ and $V_0=k\,t_{0}^{\frac{-2}{3(1+w)}}/(1+w)$. The quantum potential \eqref{qp-sw} in that case is given by
\begin{equation}\label{q-s2i}
Q(a,\phi)=-V_0\phi^{-\frac{1+3w}{3(1+w)}}-\bigg( \frac{9V_{0}^{6}}{16V_{j0}} \bigg)^{1/9}a^{-4/3}\phi^{-\frac{11+9w}{9(1+w)}}\; .
\end{equation}

\begin{figure}[H]
	\centering
	\includegraphics[width=8cm]{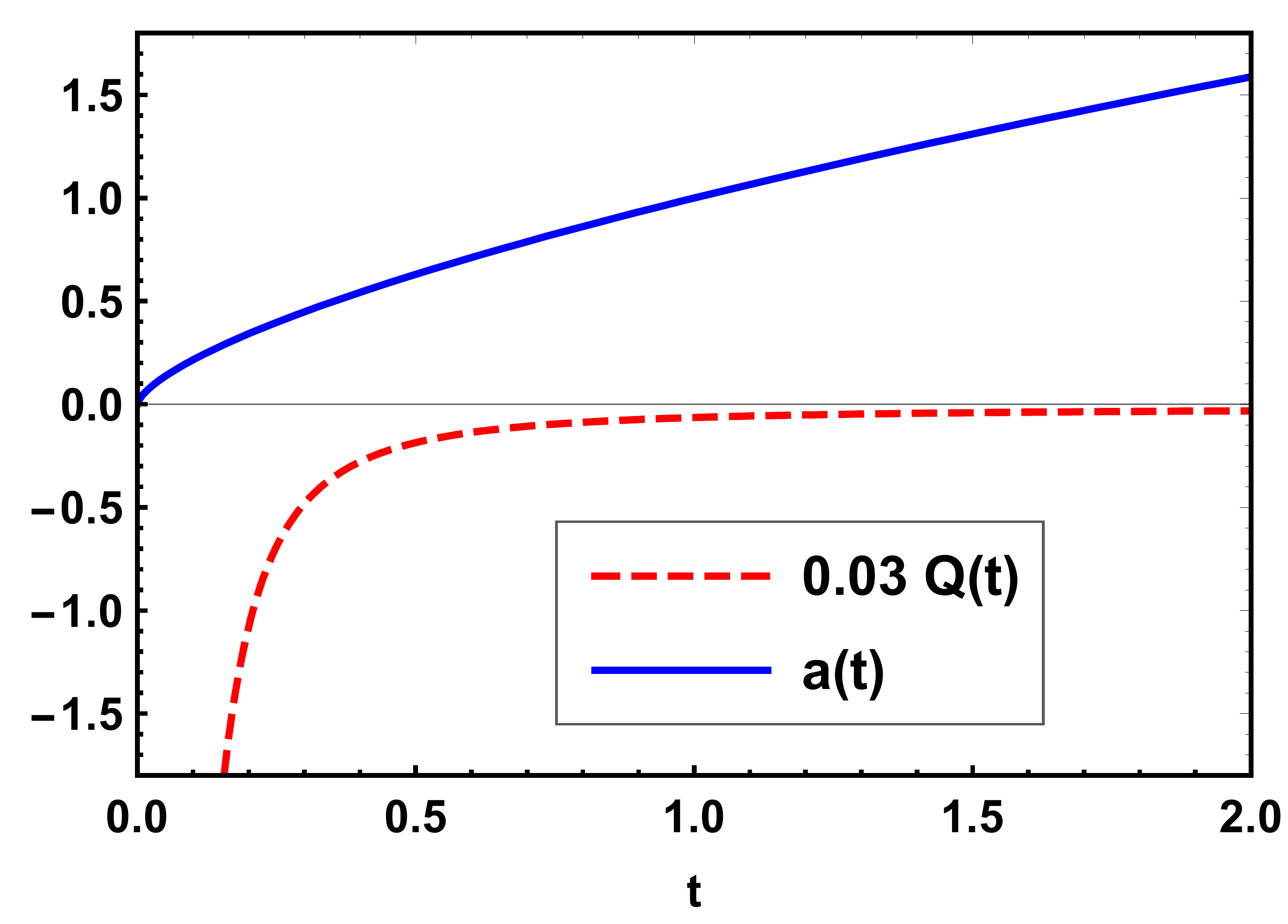}
	\caption{Time evolution of the scale factor (solid line) and the quantum potential \eqref{q-s2i} (dashed line) for \eqref{vvj-s2i}, for $k=t_0=1$, in the dust case (w=0). This solution is analogous with the perfect fluid solutions. For the other relevant cases $w=0$ and $w=1/3$, the behavior of $a(t)$ and $Q$ are very similar, as it can be seen, for instance, from the expression of $Q$ \eqref{q-s2i}.}
	\label{sol-2-i-1-dust-label}
\end{figure}

In Figure \ref{sol-2-i-1-dust-label}, we can see the time evolution of the scale factor  and the quantum potential, which diverges to $-\infty$ in $a=0$ and goes to zero when $t\longrightarrow+\infty$. We can then say that, in the present case, the quantum potential is dominant around the initial singularity $a=0$ and smooths out when the expansion dominates. Thus, the expansion takes place as the quantum potential goes to zero, meaning that the classical dynamics is recovered for this case.

\subsection{Solution (ii)}\label{subsec62}
If we choose now 
\begin{subequations}\label{vvj-s2ii}
	\begin{align}
		V_J(\phi)&=-V_{j0}\phi\,e^{9\gamma\phi}		\; ,\label{vj-62}\\
		V(\phi)&=V_0e^{\gamma\phi}		\; ,
	\end{align}
\end{subequations}
then the scale factor represents a de Sitter universe $a(t)=a_0e^{\gamma t}$, with $V_{j0}=k^3a_{0}^{9}/6$ and $V_0=3ka_0\gamma/2$. In this case, the quantum potential \eqref{qp-sw} is written as:
\begin{equation}\label{q-s2ii}
Q(a,\phi)=-V_0e^{\gamma\phi}-\bigg(\frac{9V_{0}^{6}}{16V_{j0}}\bigg)^{1/9}a^{-4/3}\phi^{-1/3}e^{-\gamma\phi/3}\; .
\end{equation}

\begin{figure}[H]
	\centering
	\includegraphics[width=8cm]{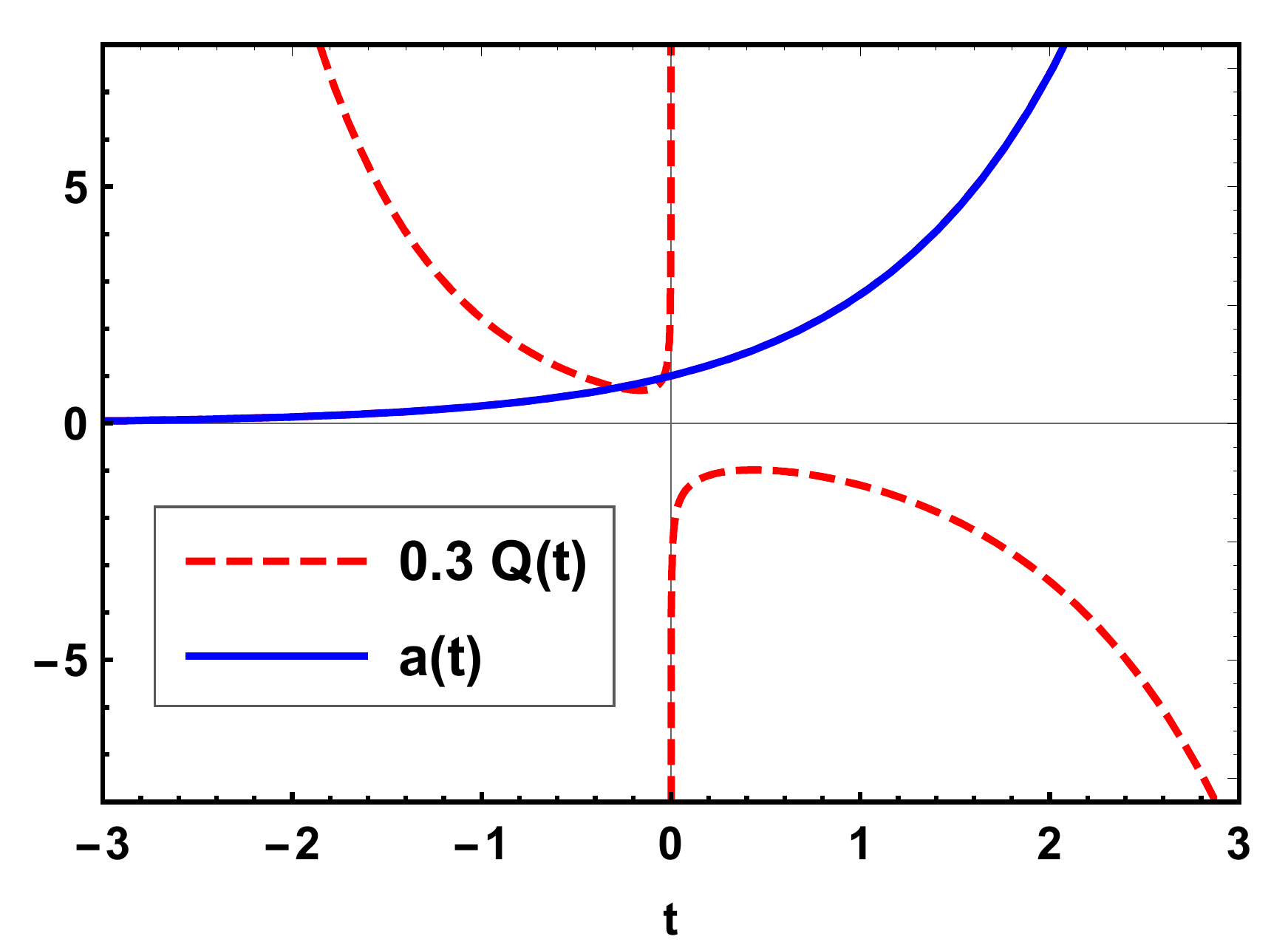}
	\caption{Time evolution of the scale factor (solid line) and the quantum potential \eqref{q-s2ii} (dashed line) for the de Sitter solution, setting $k=a_0=\gamma=1$.}
	\label{sol-2-ii-label}
\end{figure}

In Figure \ref{sol-2-ii-label}, it is shown the time evolution of \eqref{q-s2ii} and $a(t)$, where we can see that the quantum potential diverges for $t=0$ (which corresponds to $a=a_0$), but also for $t\longrightarrow\pm\infty$, and it is smooth for any $t\neq0$. Observe that the effective potential would be $Q+V$, which also diverges, in view of \eqref{q-s2ii} and Table \ref{sol-plb-class-tab}. Hence, the divergence is not a pathology from $Q$, but a feature of the energy balance itself instead, in order to generate the expansion shown in Figure \ref{sol-2-ii-label}.

\subsection{Solution (iii)}\label{subsec63}
We can also obtain the correction of Solution (i) by a bounce, $a(t)=a_0[1+(t/\phi_0)^2]^{\frac{1}{3(1+w)}}$, by choosing
\begin{subequations}\label{vvj-s2iii}
	\begin{align}
		V_J(\phi)&=-V_{j0}\phi[1+(\phi/\phi_0)^2]^{\frac{3}{1+w}}		\; ,\label{vj-63}\\
		V(\phi)&=V_0\phi[1+(\phi/\phi_0)^2]^{-\frac{2+3w}{3(1+w)}}		\; ,
	\end{align}
\end{subequations}
where $V_{j0}=k^3a_{0}^{9}/6$ and $V_0=k\,a_0/[\phi_{0}^{2}(1+w)]$. Now, the quantum potential \eqref{qp-sw} is given by
\begin{equation}\label{q-s2iii}
Q(a,\phi)=-V_0\phi\bigg[ 1+\bigg(\frac{\phi}{\phi_0}\bigg)^{2} \bigg]^{-\frac{2+3w}{3(1+w)}}- \bigg(\frac{9V_{0}^{6}\phi^3}{16V_{j0}a^{12}}\bigg)^{1/9} \bigg[ 1+\bigg(\frac{\phi}{\phi_0}\bigg)^{2} \bigg]^{-\frac{7+6w}{9(1+w)}}\; .
\end{equation}

\begin{figure}[H]
	\centering
	\includegraphics[width=8cm]{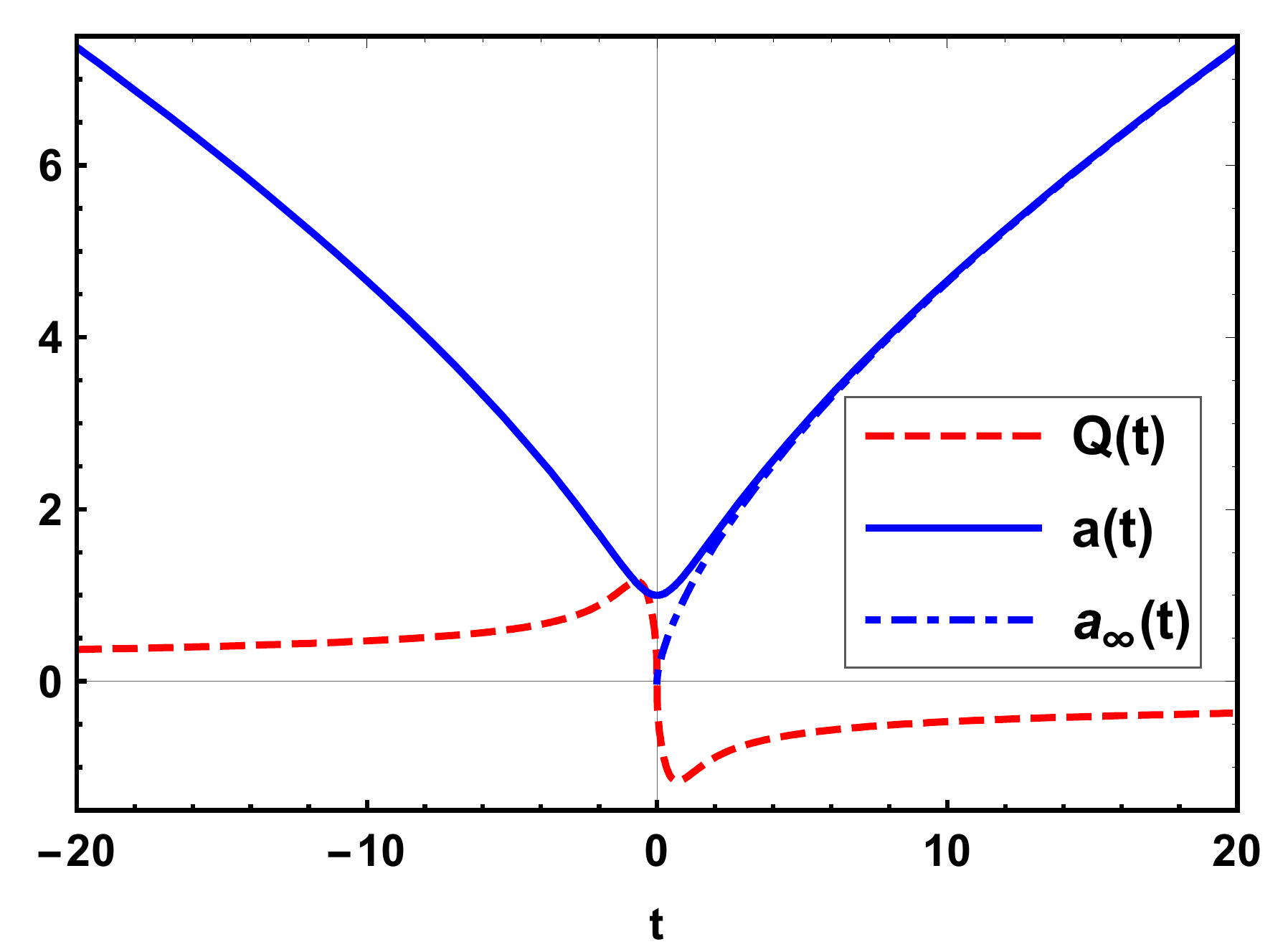}
	\caption{Time evolution of $a(t)=a_0[1+(t/\phi_0)^2]^{\frac{1}{3(1+w)}}$ (solid line) and its corresponding quantum potential \eqref{q-s2iv} (dashed line), for the dust-dominated case, with $k=a_0=t_0=\phi_0=1$. The notation $a_{\infty}(t)$ represents the asymptotic form of $a$, $a_{\infty}(t)=(t/t_0)^{\frac{2}{3(1+w)}}$ (dash-dotted line), which is equivalent to Solution (i).}
	\label{sol-2-iii-label}
\end{figure}

We show in Figure \ref{sol-2-iii-label} the time evolution of the quantum potential \eqref{q-s2iii}, for the dust case, where we can see that the quantum potential is smooth for any $t$, it dominates during the bounce, and it goes to zero for $|t|\gg0$. For the radiation-dominated and stiff matter-dominated cases, the quantum potential has exactly the same qualitative behavior, as it can be seen from \eqref{q-s2iii}. 

In mathematical terms, we can see that even though $Q\sim a^{-4/3}$, the solution $a(t)$ is non-singular, so $Q$ will never diverge and also, since there are no negative powers of $\phi$ in the expression of $Q$, it is well defined at $\phi=t=0$. In physical terms, what all those features are saying is that the quantum potential is dominant around the bounce and it goes to zero before and after the bounce, thus the avoidance of the singularity is a quantum effect generated by the correction of Hamilton-Jacobi equation provided by $Q$.

\subsection{Solution (iv)}\label{subsec64}
In analogy with the previous case, if we choose
\begin{subequations}\label{vvj-s2iv}
	\begin{align}
		V_J(\phi)&=-V_{j0}\phi\cosh^{9}(\gamma\phi)		\; ,\label{vj-64}\\
		V(\phi)&=V_0\sinh(\gamma\phi)		\; ,
	\end{align}
\end{subequations}
we obtain a bounce that avoids the de Sitter asymptotic singularity, $a(t)=a_0\cosh(\gamma t)$, where $V_{j0}=k^3a_{0}^{9}/6$ and $V_0=3ka_0\gamma/2$. For this case, the quantum potential \eqref{qp-sw} becomes:
\begin{equation}\label{q-s2iv}
Q(a,\phi)=-V_0\sinh(\gamma\phi)-\bigg[ \frac{9V_{0}^{6}\sinh^6(\gamma\phi)}{16V_{j0}a^{12}\phi^3\cosh^9(\gamma\phi)} \bigg]^{1/9}\; .
\end{equation}

\begin{figure}[H]
	\centering
	\includegraphics[width=8cm]{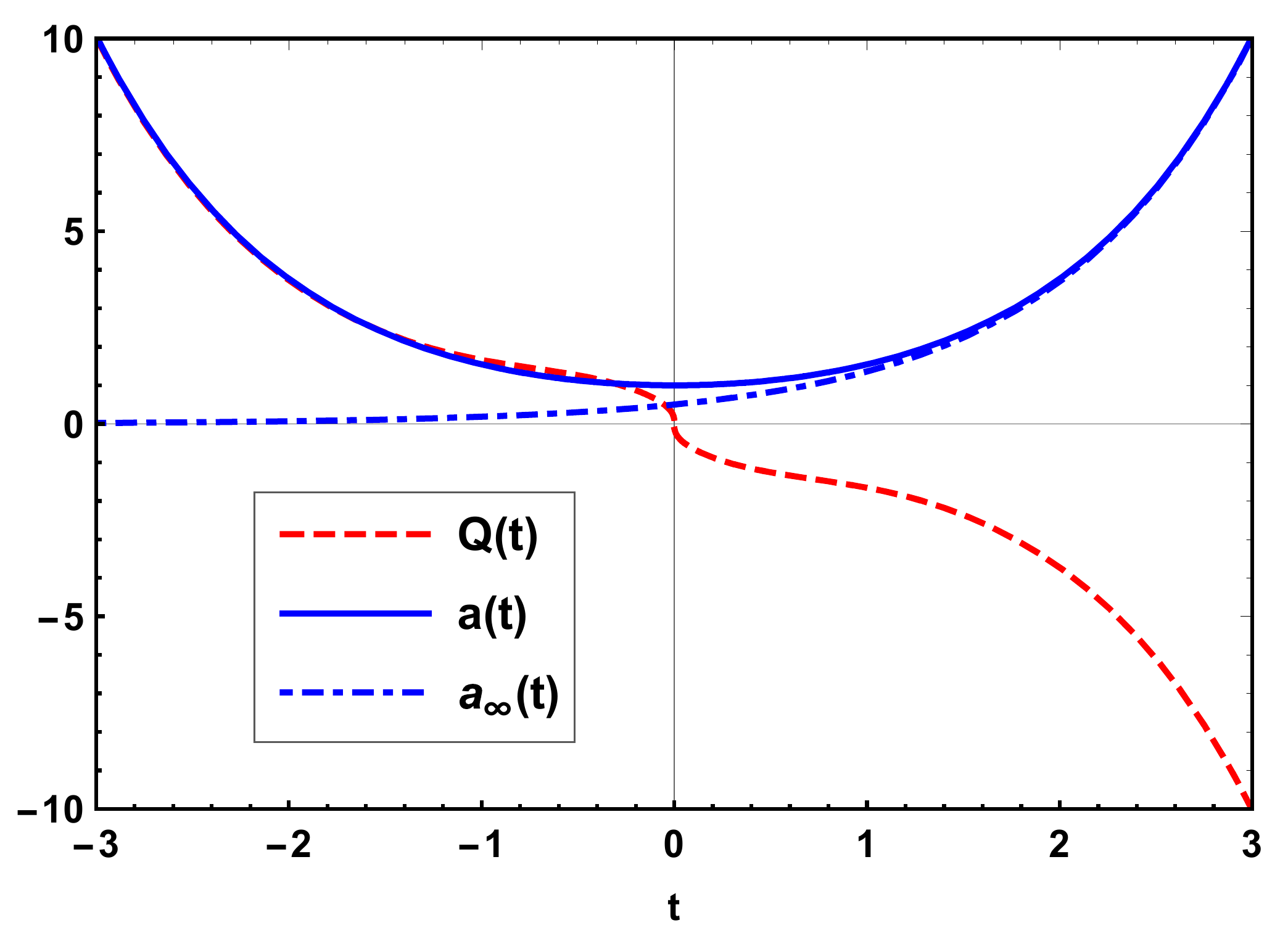}
	\caption{Time evolution of $a(t)=\cosh(\gamma t)$ (solid line) and its corresponding quantum potential \eqref{q-s2iv} (dashed line), for  $k=a_0=\gamma=1$. The notation $a_{\infty}(t)$ represents the asymptotic form of $a$, $a_{\infty}(t)=\frac{1}{2}e^{\gamma t}$ (dash-dotted line), which is equivalent to Solution (ii).}
	\label{sol-2-iv-label}
\end{figure}

That quantum potential is not well defined for $\phi=t=0$. Although, it could become a continuous function if we set $Q(0)=0$, as it can be seen evaluating the limit of \eqref{q-s2iv} when $t\longrightarrow0$. In Figure \ref{sol-2-iv-label}, we see the time evolution of the quantum potential \eqref{q-s2iv} compared with the scale factor. We can see that $Q$ diverges for $t\longrightarrow\pm\infty$. Thus, for this solution, both the contraction and the expansion are driven by $Q$.

\subsection{Solution (v)}\label{subsec65}
Finally, an oscillatory universe, for which $a(t)=a_m+A[1-\cos(\omega t)]$, is obtained choosing
\begin{subequations}\label{vvj-s2v}
	\begin{align}
		V_J(\phi)&=-V_{j0}\phi\{a_m+A[1-\cos(\omega\phi)]\}^9		\; , \label{vj-65}\\
		V(\phi)&=V_0\sin(\omega\phi)		\; ,
	\end{align}
\end{subequations}
where $V_{j0}=k^3/6$ and $V_0=3kA\omega/2$. Finally, for this case, the quantum potential \eqref{qp-sw} is:
\begin{equation}\label{q-s2v}
Q(a,\phi)=-V_0\sin(\omega\phi)-\frac{[9V_{0}^{6}\sin^6(\omega\phi)/(16V_{j0}a^{12}\phi^3)]^{1/9}}{a_m+A[1-\cos(\omega \phi)]}\; .
\end{equation}

\begin{figure}[H]
	\centering
	\includegraphics[width=8cm]{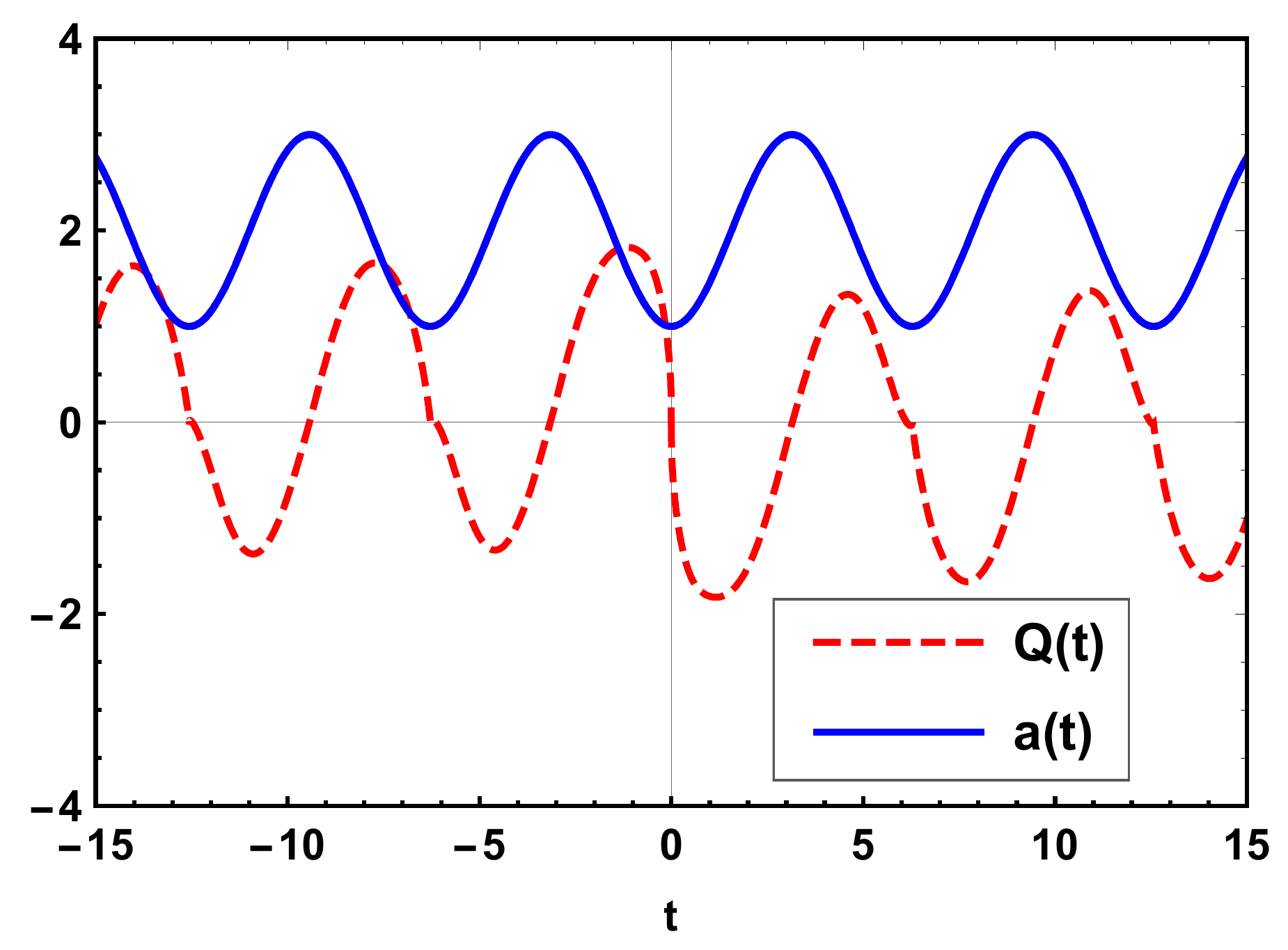}
	\caption{Time evolution of the oscillatory solution and its corresponding quantum potential  for \eqref{vvj-s2v}, with $k=A=\omega=a_m=1$.}
	\label{sol-2-v-label}
\end{figure}

In Figure \ref{sol-2-v-label}, we see the time evolution of \eqref{q-s2v} in comparison with the oscillatory scale factor. $Q$ is not well defined for $t=0$, but it can be redefined setting $Q(0)=0$ to become smooth, as it can be done for the previous bouncing solution above. When $t\neq0$, we see that $Q$ oscillates along with the scale factor and is the responsible for the oscillation.

\section{Conclusions}\label{sec-conc}
In the two last Sections, we investigated ten different possibilities to study the quantum system \eqref{guidk}. We now make some last comments about the obtained results. Since the term that governs the dynamics is the quantum potential \eqref{qp-sw}, we will focus on the relation between $Q$ and the scale factor, which gives us a criterion to interpret and classify all the solutions presented here. The quantum solutions from Sections \ref{sec-sol1} and \ref{sec-sol2} can be organized in three classes.

The first class includes all solutions from Section \ref{sec-sol1} and also the ones from Subsections \ref{subsec61} and \ref{subsec62}. In all those cases, the quantum potential is smooth almost everywhere, but it diverges to $\pm\infty$ at some point. That can be understood in two different ways. It is as a problem if we compare it with a more common case, like minimal coupling, where the quantum potential is usually small and don't have divergences, at least for the majority of cases. But this interpretation is not necessarily valid, since there are also canonical examples of real physical systems for which the quantum potential diverges. This is the case, for instance, for the ground state of hydrogen-like atoms in the Bohm-de Broglie interpretation of Schr\"odinger equation, which is in exact accordance with the numerical results of the standard interpretations \cite{holland_1993}. If we assume the latter understanding, the divergence tells us that the quantum effects are very strong around the particular divergence point, or asymptotically. Take, for instance, Figure \ref{sol-i-iv-label}. There we can see that the quantum potential diverges for $t=0$, which corresponds to $a=0$, and since the solution is singular, the divergence of $Q$ at $t=0$ is related with the singularity $a=0$ at the same time. And, for the another singular point, the quantum potential shows to be relevant again.

The solution from section \ref{subsec61} is in the first class, but it deserves a specific interpretation, because even though its quantum potential diverges at the singularity $a=0$, it goes to zero as time pass, thus indicating the recovery of classical dynamics when the universe expansion dominates. For the other cases in the first class of solutions, the fact that the quantum potential dominates even when the universe is expanding means that the expansion is a quantum effect for those solutions. In summary, we can say that the first class of solutions features an unusual behavior for the quantum potential, which is, nevertheless, not actually forbidden in a Bohm-de Broglie setting.

The second class of quantum solutions corresponds to the ones from Subsections \ref{subsec64} and \ref{subsec65}. For both, $Q$ is not well defined at $t=0$, because we restricted the discussion to the case $\phi=t$ and in both cases $Q$ is proportional to some negative power of $\phi$. Thus, strictly speaking, those solutions are problematic, because the quantum potential is not defined for an instant in time that do not corresponds to a singularity in $a$, since the scale factors from Subsections \ref{subsec64} and \ref{subsec65} are both non-singular. It may be possible to avoid that problem by interpreting the scalar field in a different manner for those solutions, which can be investigated in future works.

The quantum solutions from Subsection \ref{subsec63} constitute the third class. The quantum potential \eqref{q-s2iii} is smooth, well defined for all times and it goes to zero as $a(t)$ increases (see Figure \ref{sol-2-iii-label}), experimenting only a tiny oscillation around $t=0$. That behavior means that $Q$ is more relevant around $t=0$, when the bounce happens, and $Q$ is smaller for the other values of time, when the scale factor is larger. In other words, that bounce can be considered as a quantum effect caused by $Q$ and the classical world is recovered in large scales, since a decreasing $Q$ makes the universe go back to the classical Hamilton-Jacobi dynamics of \eqref{hjclass}. Now, taking into account that the formalism with canonical transformations from \cite{TORRES2019135003} was not able to deal with the quantization of solution $a(t)=a_0[1+(t/\phi_0)^2]^{\frac{1}{3(1+w)}}$, in opposition with the fractional calculus method, we can say that the formalism developed here actually made a contribution to the quantization of solutions from \cite{TORRES2019135003}. In physical terms, it means that the bounce which approaches the perfect fluid solutions when the universe expands received a quantum justification, in accordance with what is expected from a quantum formalism with Bohm-de Broglie interpretation.

An important question about quantum bouncing solutions analogous to perfect fluid ones is the matter density during the bounce, as it was pointed out in \cite{PhysRevD.74.084003} and references therein. In some situations, it may happen that the matter density is too small during the bounce, which would be problematic, since the density must decrease with expansion. In the present quantum theory, there is no clear way how we should define the matter density, except for the solutions analogous to perfect fluid ones, which were studied in Subsections \ref{subsec61} e \ref{subsec63}. Let us then consider the bouncing case, which is the solution from \ref{subsec63}: $a(t)=a_0[1+(t/t_0)^2]^{\frac{1}{3(1+w)}}$, where $t_0$ is an integration constant. The  continuity equation implies that $\rho=\rho_0 a^{-3(1+w)}$, where $\rho_0$ is constant. If  $t=0$ represents the moment when the bounce happens, then the scale factor of the bounce is $a_0=a(t=0)$. Representing the time interval between the bounce and the present time by $T$, the matter density at the bounce is given by $\rho_B=\rho_0a_{0}^{-3(1+w)}$, whilst the matter density today is 
\begin{equation}
\rho_T=\frac{\rho_0a_{0}^{-3(1+w)}}{1+(T/t_0)^2}\;.
\end{equation}
This implies that 
\begin{equation}
\frac{\rho_B}{\rho_T}=1+(T/t_0)^2\;,
\end{equation}
from which we can conclude that the matter density at the bounce was very high compared with present day matter density, since the asymptotic condition is $T/t_0\gg 1$. And the precise value of $t_0$, which gives the asymptotic limit, is related to the quantization, since $t_0$ depends on the value of the quantum number $k$ by $t_0=ka_0/V_0(1+w)$. In summary, the bounce density is related to the quantization and from that we can infer that the problem of the small bounce density is not present for that solution.

As a last comment, we can say that the fractional formalism developed here has shown to be a valuable tool for quantum cosmological theories, in the sense that it has provided a mechanism to study several alternative cosmological scenarios. There are, of course, some solutions that need further studies and some questions remain open for future works. The application of CFD has proven to provide a handful approach to the Wheeler-DeWitt quantization of Hamiltonians like \eqref{ham}, which has momenta with fractional powers, thus making it possible to extend this fractional method to other theories, like the k-essence studied in \cite{10.1007/978-3-319-69164-0_25}. Another natural extension of the method developed here is the application of other fractional derivatives, which may also reveal complementary points of view for the quantum effects.\\

\vspace{12pt}
\noindent {\bf Acknowledgments}

We thank to Felipe de Melo Santos, Felipe Tovar Falciano, Ingrid Ferreira da Costa, Jorge Zanelli, Joseph Buchbinder, and Nelson Pinto-Neto for very important discussions about this paper. This study was financed in part by the \emph{Coordena\c{c}\~ao de Aperfei\c{c}oamento de Pessoal de N\'ivel Superior} - Brazil (CAPES) - Finance Code 001 and also by FAPES and CNPq from Brazil. OFP thanks the Alexander von Humboldt foundation for funding and the Institute for Theoretical Physics of the Heidelberg University for kind hospitality.

\bibliographystyle{unsrt}
\bibliography{references}

\end{document}